\documentclass[twocolumn,runningheads]{svjour2}
\journalname{Astrophysics and Space Science}
\usepackage{graphicx}
\usepackage[usenames]{color}
\usepackage{multirow}
\begin{document}

\title{Equation of state   of neutron star cores and spin down of isolated
pulsars}
\author{P. Haensel
\and
 J.L.Zdunik
 }
\institute{N. Copernicus Astronomical Center, Polish
           Academy of Sciences, Bartycka 18, PL-00-716 Warszawa,
           Poland\\
           \email{haensel@camk.edu.pl, jlz@camk.edu.pl}}

\titlerunning{EOS and spin down of pulsars}
\authorrunning{P. Haensel and J.L. Zdunik}

\date{}

\maketitle

\begin{abstract}
We study possible  impact   of a softening of the
equation of state by a phase transition, or appearance of
hyperons, on the spin evolution of of isolated pulsars.
Numerical simulations are performed using exact 2-D simulations in general
relativity. The equation of state of dense matter at supranuclear
densities is poorly known. Therefore, the accent is put on the
general correlations between evolution and equation of state,
 and mathematical strictness.
General conjectures referring to
 the structure of the one-parameter families
of stationary configurations are formulated. The interplay
of the back bending
phenomenon and stability with respect to axisymmetric perturbations
is described. Changes of pulsar parameters in a corequake following
instability are discussed, for a broad choice of phase transitions
predicted by different theories of dense matter.  The energy release in
a corequake, at a given initial pressure,  is shown to be independent
of the angular momentum of collapsing
configuration.  This result holds  for various types of phases transition,
 with and without metastability. We critically review
 observations of pulsars that could be relevant for the
 detection of the signatures of the phase transition in
 neutron star cores.

\end{abstract}

%
\section{Introduction}
\label{sec.introd}
The equation of state (EOS) above a few times nuclear density is a
main mystery of neutron stars. Our degree of ignorance
concerning EOS of neutron star core is therefore high. In particular,
different theories of dense matter predict various types of softening of the
EOS above some density, due to phase transition to exotic phases of
hadronic matter (meson condensates, quark matter), or presence of new baryons,
such as hyperons (for a review, see Glendenning 2000).

Since their discovery in 1967, thanks to the exceptional precision
of their timing, isolated radio pulsars are cosmic laboratories
to study physics of ultradense matter. Such a pulsar looses its angular momentum, $J$,  due
to radiation. In response to the angular momentum loss, $\dot{J}<0$,
the pulsar  changes its
rotation frequency  $f=1/{\rm period}=\Omega/2\pi$ (measured
in pulsar timing) and increases its central density and
pressure $\rho_{\rm c}, ~P_{\rm c}$. The response of the stellar
structure to $\dot{J}$ depends on the equation of state (EOS)
of the neutron star core,
and is sensitive to appearance of new particles or of a new phase at the
neutron star center. In particular, crossing the
phase-transition region by $\rho_{\rm c}$ is reflected by
specific ``nonstandard'' behavior of $f$ as a function of time.
In the present
paper we analyze the relation  between the pulsar spin down
and the EOS. This problem was studied previously by many
authors, who considered various types of EOSs of dense matter
 predicted by different  theories
 (Weber \& Glendenning 1991, 1992; Glendenning, Pei, \& Weber 1997,
Heiselberg \& Hjorth-Jensen 1998, Chubarian et al. 2000, Spyrou \& Stergioulas 2002).
In the present paper we summarize and review results obtained
using a different approach. Namely, through a flexible
parametrization of dense matter EOS,  we are able study the problem of
spin down evolution, including stability and the back bending
phenomenon (spin up accompanying angular momentum loss),
for a very broad, continuous family of EOSs with and
without phase transitions, and select those features which are
genuine properties of the EOSs. On the other hand, the use of
parametrized EOSs allows  to perform very  precise 2-D
calculations, crucial for reliable analysis of stability, and
avoiding  usual severe limitations resulting from the lack of
precision.

The plan of the present paper is as follows. In
Sect.\ \ref{sect:EOSequil} we
review effect of phase transitions on the the equation of
state (EOS) of dense matter under the assumption of  full
thermodynamic equilibrium. Then in Sect.\ \ref{sect:stab} we
remind stability criteria for relativistic rotating stars.
Importance of stability for the back bending phenomenon is
discussed in Sect.\ \ref{sect:BBstab}.
In Sect.\ \ref{sect:invar} various cases of evolutionary tracks
of isolated pulsars, corresponding to different EOSs of
dense matter, are studied, and general properties of the
structure of stable families of stationary configurations
are deduced.  Corequakes resulting from
instability of rotating configurations are discussed in
Sect.\ \ref{sect:InstabQuake}. In Sect.\
\ref{sect:BBtimingPSR} we show how  a phase transition in
neutron star  core can affect the time evolution of the pulsar
rotation period.
Some remarkable features of changes in pulsar parameters resulting from
 a corequake  are discussed in
Sects.\ \ref{sect:DeltaQuakes}-\ref{sect:MetastabQuakes}.  Finally, Sect.\
\ref{sect:Conclusions} contains discussion and conclusions.

Results reviewed here were  obtained in last few years within a fruitful
collaboration with M. Bejger and E. Gourgoulhon (Zdunik et al.
2004, 2006a, 2006b, Bejger et al. 2005). We present these results in  the
most general terms, and summarize them in the form of
conjectures  (which in the future may become strict
theorems), which on the one hand are general
(i.e., do not depend on a specific dense matter model), and on
the other hand seem to us useful for further numerical studies
of the intimate relation  between EOS and neutron star dynamics.
\section{First order phase transition in thermodynamic equilibrium }
\label{sect:EOSequil}
\begin{figure}
\includegraphics[width=7cm,draft=false]{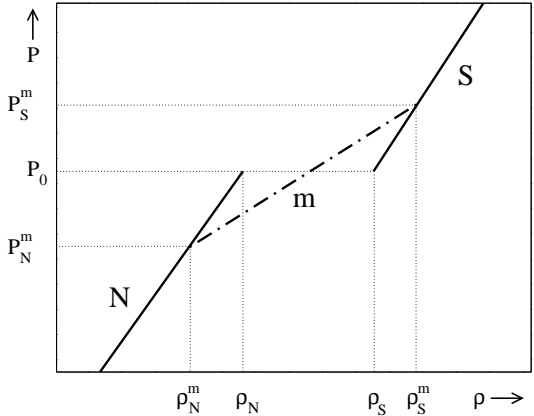}
\caption{Two types of the EOS with 1st order phase transition.
Thick solid lines: stable pure N and S phases. Mixed phase is
represented by a dash-dot line. For further
explanation see the text. From Bejger et al. (2005)}
\label{fig:EOS-equil}
\end{figure}

Two possible cases of the EOS softening by a 1st order phase transition
are  shown schematically in Fig.\ \ref{fig:EOS-equil}. Simplest case is
that of a first order phase transition from a pure N (normal)
phase to a pure S (superdense) phase. It occurs at a well
defined pressure $P_0$, and is accompanied by a  density
jump: $\rho_{\rm N}<\rho_{\rm S}$. Such a type of phase
transition occurs for sufficiently strong pion or kaon
condensation, and is also characteristic for many models of
quark deconfinement.
For a sufficiently small  surface tension at the N-S interface,
$\sigma<\sigma_{\rm crit}$, one has to contemplate a more
complicated pattern of the N-S phase coexistence. Above some
$P_{\rm N}^{\rm m}$  a mixture NS is preferred  over the
pure N phase, with fraction of the S phase increasing from
zero to one at $P=P_{\rm S}^{\rm m}$, and above $P_{\rm
S}^{\rm m}$ pure S phase is preferred. Such a situation
might be possible for meson (kaon or pion) condensations or
quark matter, provided $\sigma<\sigma_{\rm crit}$.
Alas, our knowledge
of physics of superdense matter is not sufficient to decide
which case (pure or mixed) type of first order phase
transition is actually realized in dense neutron star cores.
\section{ Stability of hydrostatic equilibria}
\label{sect:stab}
Neutron stars are assumed to be built of an ideal fluid. Therefore,
 stresses resulting from elastic shear, electromagnetic field,
 viscosity, etc. are neglected. \par
In our case, stability criteria are:
\vskip 2mm
\parindent 0pt
{\it Stability criteria - nonrotating stars}
Nonrotating configuration form a one-parameter family ${\cal
C}(x)$, and are labeled by  $x=\rho_{\rm c}$ or $x=P_{\rm c}$.
One considers stability of these 1-D configurations with
respect to radial perturbations and finds that a configuration
is stable if  ${\rm d}M/{\rm d} x
>0$ and unstable if ${\rm d}M/{\rm d} x <0$ (Harrison et al.
1966).
\vskip 3mm \parindent=0pt
 {\it Stability criteria - rotating
stars}. They were precisely formulated by Friedman et al. (1988).
 Stationary rotating configurations form a two-parameter family:
 ${\cal C}(x,\Omega)$. Their stability with respect to
  axially symmetric
perturbations can be checked using any of the following criteria,
involving baryon (rest) mass $M_{\rm b}$ or  stellar
angular momentum $J$ :\par
\vskip 3mm
\parindent 0pt
 {\bf (1)} configuration is
 stable if $(\partial M/\partial x)_{J=const.}>0$ and
unstable if
 $(\partial M/\partial x)_{J=const.}<0$~,\par\vskip 2mm
 ~~~~~~or\par
 \vskip 2mm
{\bf (2)} configuration is
stable if  $(\partial J/\partial x)_{M_{\rm
b}=const.}<0$ and unstable if
 $(\partial J/\partial x)_{M_{\rm b}=const.}>0~.$
\section{Back-bending and stability of rotating configurations}
\label{sect:BBstab}


\begin{figure}
\includegraphics[width=7.5cm,draft=false]{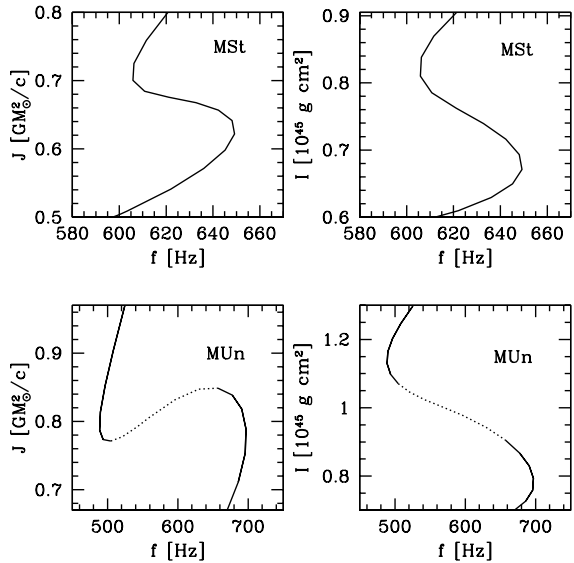}
\caption{Total angular momentum  versus rotation frequency $f$ (left panels),
 and moment of inertia $I\equiv J/\Omega$ versus $f$ (right panels),
  for EOSs  MSt and  MUn of Zdunik et al. (2006a).
The  stability criterion is easily applied to left panels. It
is clear  that for the MSt EOS back bending feature is not
associated with an instability, with all configurations being
stable. On the contrary, the  MUn EOS produces back bending
with a large segment of unstable configurations.
Simultaneously,  the $I(f)$ curves for both EOSs are very
similar and apparently show  very  similar back-bending
shapes. }
 \label{ji}
\end{figure}
%
Zdunik et al. (2006a) have shown examples of the EOS with
 1st order phase transitions between the pure N and S phases,  as well
 as  mixed-phase  transitions,  for which one of the two situations
 occurs:

\begin{itemize}
\item{1. all configurations are stable but the back bending
exists.}

\item{2. the phase transition results in the instability region for rotating
stars.}
\end{itemize}

To distinguish between these two cases we can look at the behavior
either of the curves $M_{\rm b}(\rho_{\rm c})_{J}$ or
$J(f)_{M_{\rm b}}$; the
quantity fixed along a sequence is indicated by the lower index.
For such curves  the instability criterion directly applies.
However, is not so easy to detect the instability using the
$I(f)_{M_{\rm b}}$ plot, where $I\equiv J/\Omega$ is stellar moment of inertia.
The unstable case may look similar to that corresponding to a
fully stable sequence. An  example is presented in Fig.\ \ref{ji},
 where we plot two functions:
$J(f)_{M_{\rm b}}$  and $I(f)_{M_{\rm b}}$ for the two models of EOS
 (MSt and MUn) with softening due a mixed-phase segment.
 Upper panels correspond to the stable model MSt and lower to the
MUn  model for which exist region of unstable configurations. The
difference between the MSt and MUn cases is clearly visible in
left panels ($J(f)_{M_{\rm b}}$ curves), where we can easily recognize
the instability region for MUn model, by  applying the condition
${({\rm d}J/{\rm d}\rho_{\rm c})}_{M_{\rm b}}>0$, and keeping in mind that
$\rho_{\rm c}$ is monotonic along this curve. However, the upper and
lower right panels ($I(f)_{M_{\rm b}}$) are quite similar,  without any
qualitative difference. Simultaneously,  the  back bending phenomenon in the
MUn case is very large (almost by 200~Hz),  and obviously the
phase transition results in an instability (increase of $J$ for
increasing $\rho_{\rm c}$). An yet,  $I(\rho_{\rm c})$ is  a monotonic
decreasing function all the time.
\section{ Invariance of structure of (one-parameter)
 families $\lbrace{\cal C}_X\rbrace$}
 \label{sect:invar}
\begin{figure}
\includegraphics[width=6.5cm,draft=false,angle=-90]{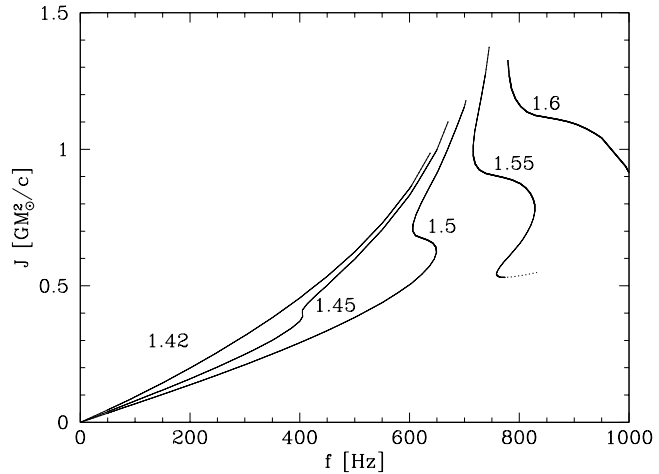}
\caption{Examples of the evolutionary tracks in the
$J-f$ plane for an EOS which produces back bending but  does
not lead to splitting of the tracks into disjoint stable
branches, separated by unstable segments.
Curves are labeled by baryon mass (constant along a track)
 in solar masses.
Thick segments correspond to back bending phenomenon. A small
dotted termination  at the lower end of the 1.55 curve
corresponds to a collapse into black hole after reaching the
minimum $J$ allowed for stationary configurations.
From Zdunik et al. (2006a).
} \label{fig:StableMRJf}
\end{figure}

\begin{figure}
\includegraphics[width=6.5cm,draft=false,angle=-90]{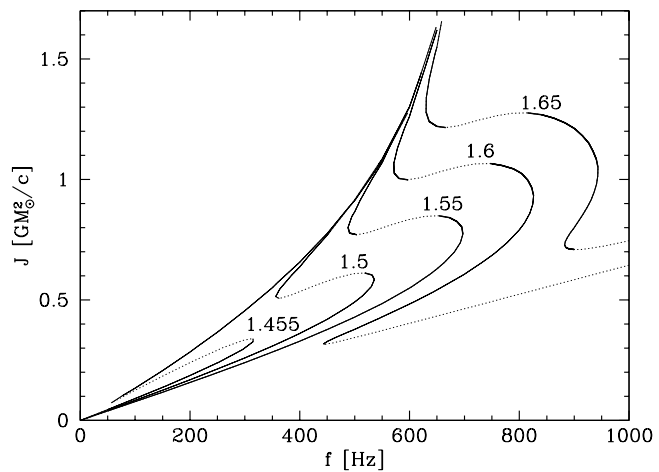}
\caption{Examples of the evolutionary tracks in the
$J-f$ plane for an EOS which produces unstable segments.
Curves are labeled by baryon mass (constant along a track)
 in solar masses.
Thick segments correspond to back bending phenomenon.
From Zdunik et al. (2006a).
} \label{fig:UnstMRJF}
\end{figure}

 \parindent 21pt
We select a one-parameter family (represented by a curve)
from a two-parameter set
of stationary configuration by fixing one of parameters, denoted
by $X$, so that within this family $X=const.$
Here $X=M_{\rm b},J,f$.  Examples of such one-parameter
families are shown in Fig.\ \ref{fig:StableMRJf},
\ref{fig:UnstMRJF}. As shown by Zdunik et al. (2006a),
studying stability of these one parameter families reveals a
very interesting invariance property.
This property can be generalized to include also the special
case of marginally unstable configurations, which correspond
to the inflection point
 $(\partial M/\partial x)_{J=const.}=0$ and
  $(\partial^2 M/\partial x^2)_{J=const.}=0$.
For the $M_{\rm b}=const.$ families the marginal instability
corresponds to $(\partial J/\partial x)_{M_{\rm b}=const.}=0$
and $(\partial^2 J/\partial x^2)_{M_{\rm b}=const.}=0$.

Only stable configurations are astrophysically relevant.
In what follows, we use the term  ``unstable segment'' of one parameter
family in the restricted sense: we will always mean a segment composed of
unstable configurations which is bounded on both sides by stable
configurations.  A remarkably general  invariance property
 can be formulated as three conjectures:
\par\vskip 3mm
\parindent 0pt
 {\bf (1)} All stable $\lbrace {\cal C}\rbrace_{\rm
stat}$ remains all stable $\lbrace {\cal
C}_X\rbrace_{\rm rot}$\par\vskip 2mm
{\bf (2)} If $\lbrace {\cal
C}\rbrace_{\rm stat}$ contains unstable segment then every
$\lbrace {\cal C}_X\rbrace_{\rm rot}$ contains unstable segment
too\par\vskip 2mm
{\bf (3)} If $\lbrace {\cal C}\rbrace_{\rm stat}$ contains a
marginally unstable ${\cal C}$ then  each $\lbrace {\cal
C}_X\rbrace_{\rm rot}$ contains a marginally unstable ${\cal C}_X$ \par
\vskip 3mm
\parindent 21pt
These conjecture are based on very precise numerical
2-D simulations performed for hundreds of EOSs with
softening due to phase transitions. They can also be rephrased
in a more compact form:
\parindent 0pt
\vskip 2mm
${\bf 1^\prime}$ single family of stable static configurations
 $\Leftrightarrow$ single family of stable rotating  configurations
\vskip 2mm
 ${\bf 2^\prime}$ two disjoint  families of stable static
configurations separated by a family of unstable
configurations $\Leftrightarrow$
 two disjoint  families of stable rotating configurations
separated by a family of unstable configurations
 (constant $M_{\rm b}$,
 or constant $J$,  or
constant $f$)\par
\vskip 2mm
\parindent 21pt
Therefore, topology of the one-parameter families of hydrostatic
stationary equilibrium configurations is a genuine feature of
the EOS, not altered by rigid rotation.
\section{Instabilities and starquakes}
\label{sect:InstabQuake}
\begin{figure}
\includegraphics[width=\hsize,draft=false]{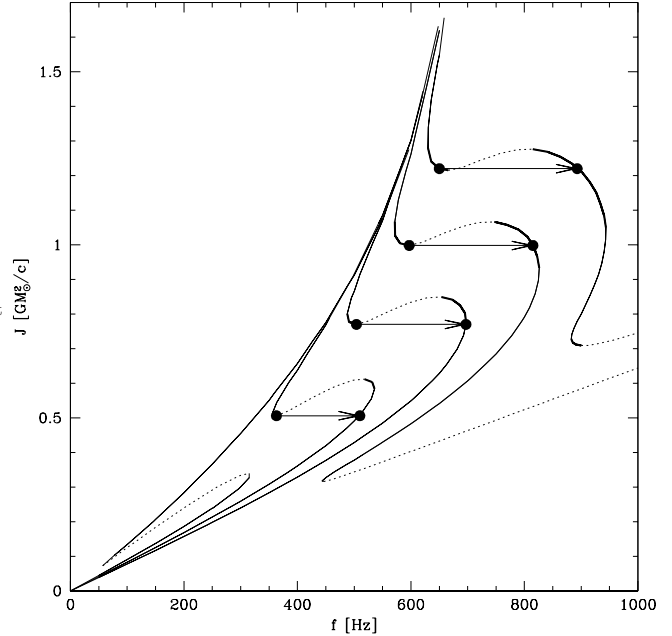}
\caption{Evolutionary tracks of an isolated pulsar loosing $J$
in the $J-f$ plane, when a softening of the EOS
due to a phase transition imply instability regions (dotted segments).
Arrows lead
from unstable configuration to a stable collapsed one, with the same
baryon mass and angular momentum. From Zdunik et al. (2006a).
 }\label{fig:BBquake}
\end{figure}

\begin{figure}
\includegraphics[width=6.5cm,draft=false,angle=-90]{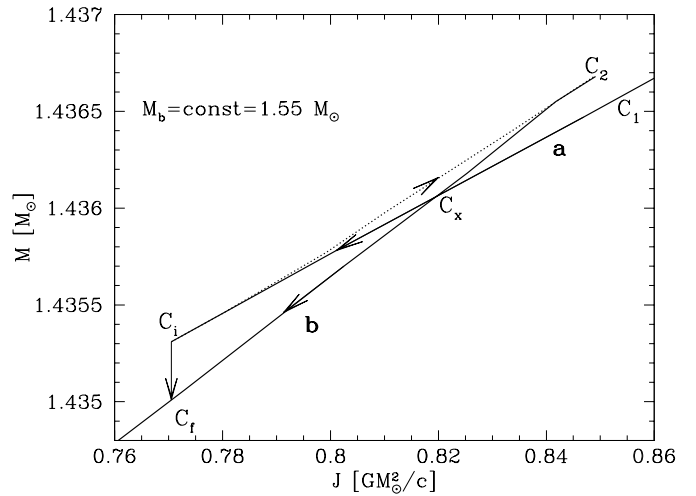}
\caption{Gravitational mass of the star as a function of its
angular momentum, for a fixed baryon mass of the star. Central
density is increasing along the curve as indicated by the arrows.
The upper (dotted) segment corresponds to the unstable
configurations. From Zdunik et al.(2006a). }\label{fig:TracksMJ}
\end{figure}

Entering the unstable segment of the evolutionary track, during pulsar
spin down, leads to a discontinuous transition (starquake) to a
stationary configuration on another stable segment of the evolutionary track.
This is illustrated by examples of evolutionary tracks in the $J-f$ plane in
Fig.\ \ref{fig:BBquake}.  It is  assumed that there is no matter ejection
during the quake and that the timescale of it is so short that the angular
momentum loss (by radiation of electromagnetic and gravitational waves)
can be neglected. Therefore, a spinning down pulsar reaches instability point
and collapses (with spinning up!)  into a new stable configuration
 ${\cal C}$ with the same baryon mass and angular momentum, and  then continues its evolution
moving in the $J-f$ plane (Fig.\ \ref{fig:BBquake}).
The unavoidability of a corequake is best seen by showing the
track in the $M-J$ plane in Fig.\ \ref{fig:TracksMJ}.
 The only way out from ${\cal
C}_{\rm i}$ is collapse to ${\cal C}_{\rm f}$, keeping $M_{\rm
b}$ and $J$ constant. The energy released during the corequake
is $\Delta E=(M_{\rm i}-M_{\rm f})c^2$. Let us stress, that
the corequake is accompanied by differential rotation,
redistribution of the angular momentum, and breaking of the axial
symmetry. It is a nonstationary phenomenon which is to be
studied using methods beyond those applied to describe 2-D rigid rotation
in the present paper.
\section{Back bending and pulsar timing}
\label{sect:BBtimingPSR}
\begin{figure}
\includegraphics[width=\hsize,draft=false]{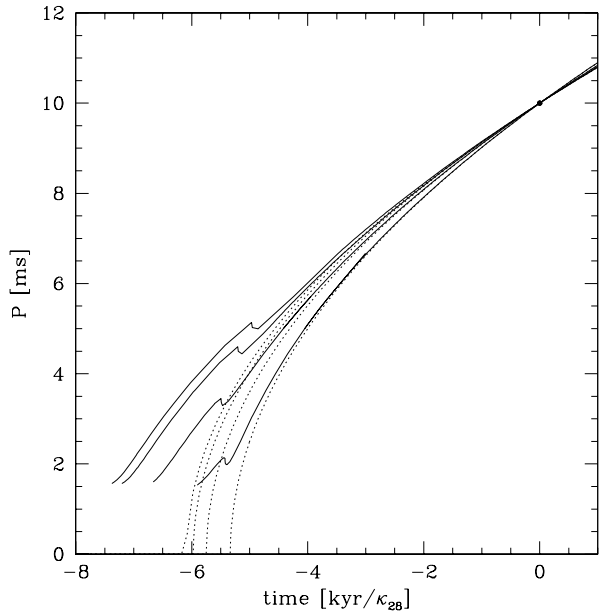}
\caption{The evolution of the pulsar period $P$ when the
energy loss is described by the magnetic dipole breaking with
$\alpha=4$. Solid curves - results obtained for an EOS with a
phase transition, for different baryon masses. Dotted lines
correspond to a standard model with $I=I(0)=const.$
The unit of time is $1000~{\rm yrs}/\kappa_{28}$, where
$\kappa_{28}=\kappa/10^{28}$ (in the c.g.s.  units), and $\kappa$
is a parameter in the  right-hand-side of Eq.\ (\ref{eq:brake}). From Zdunik et
al. (2006a).
}\label{fig:PulsarP-t.BB}
\end{figure}

A spinning down isolated pulsar looses its angular momentum and energy.
The energy balance is described by the standard formula
\begin{equation}
\dot{M}c^2=-\kappa
\Omega^\alpha~,
\label{eq:brake}
\end{equation}
where the right-hand-side describes the energy loss due to
radiation.
The standard assumption is that the $\Omega$-dependence of the
moment of inertia $I$ can be neglected, so that during spin down
 $I=I(\Omega=0)=const.$ However, in the phase transition epoch
 the $\Omega$ dependence of $I$ is crucial for the evolution
 track, timing,  and stability. Standard approximation
 $I=I(\Omega=0)=const.$ leads to a severe  {\it overestimate} of the
  spindown rate and an {\it underestimate} of pulsar  age.
An example visualizing this effect is shown in Fig.\ \ref{fig:PulsarP-t.BB}.
 We assume that a
  $P=10$ ms pulsar is observed at  time $=0$. As we see,
  evolution back in
time using $I=const.$ (dotted lines, corresponding to different
$M_{\rm b}$) can be very misleading if a phase transition
region was crossed in the past. Let us notice that in that
``phase transition epoch'' the breaking index
$n=\Omega\ddot{\Omega}/\dot{\Omega}^2$ was reaching huge
values. Breaking index $\gg 1$ is a signature of a phase
transition taking place at the pulsar center.
\section{Spin up, shrinking of radius, and energy release in
 starquakes in pulsars}
 \label{sect:DeltaQuakes}
 \parindent 21pt
The changes in $f$, equatorial radius $R_{\rm eq}$, and the
energy release can be evaluated by comparing the relevant
quantities in the stationary,
rigidly rotating states (initial and final),
${\cal C}_{\rm i}\longrightarrow {\cal C}_{\rm f}$. Assumed
conditions are  $M_{\rm b,i}=M_{\rm b,f}$, $J_{i}=J_{\rm f}$.
The energy release is then given by $\Delta E=-\Delta M c^2$.

\begin{figure}
\includegraphics[width=7.5cm,draft=false]{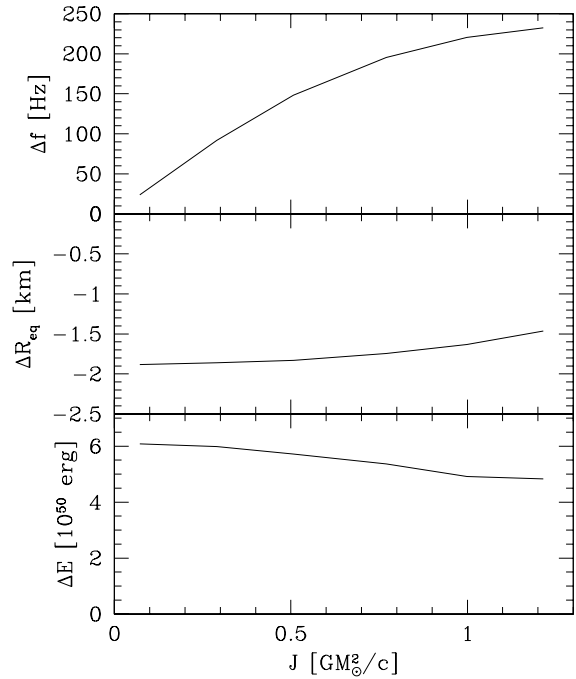}
\caption{Changes of stellar parameters of a rotating isolated
neutron stars, due to a corequake which occurs after the star reaches
an unstable configuration. An example from Zdunik et al. (2006a). }
\label{fig:DeltafRE}
\end{figure}

Examples of numerical results obtained in 2-D simulations
are shown in Fig.\ \ref{fig:DeltafRE}.  A remarkable feature is a
very weak dependence of $\Delta E$ on $J_i$.
Therefore $\Delta E$ can be calculated using 1-D code for non-rotating stars
and this gives excellent prediction (within better than 20\%) even
for high $J_{\rm i}$, when the stars are strongly flattened by rotation!
This has important practical consequence: there is no need for 2-D
simulations  to get reliable estimate estimate of the energy
release, 1-D calculation for non-rotating stars is sufficient.
\section{Metastability and two types of corequakes}
\label{sect:MetastabQuakes}
\parindent 21pt
The case of spherical, non rotating neutron stars was studied
long time ago (Haensel et al. 1986, Zdunik et al. 1987).
Corequakes in rotating neutron stars were recently studied by
Zdunik et al. (2006a,b).

\begin{figure}
\includegraphics[width=8.0cm,draft=false]{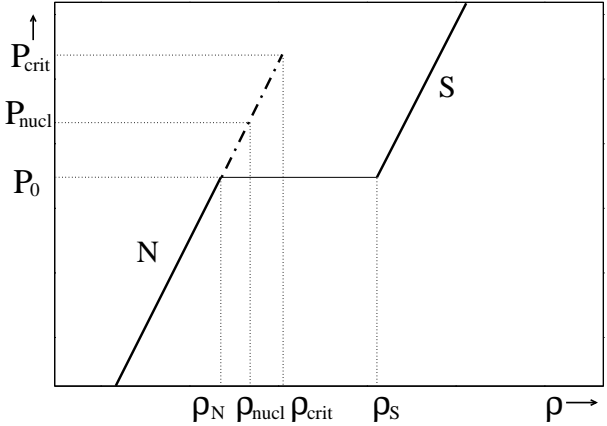}
\caption{A schematic representation, in the $\rho-P$ plane,
of an EOS with a 1st order phase transition. Solid segments: stable
 N and S phases in thermodynamic equilibrium. Dash-dot segment:
  metastable N phase. The S phase nucleates at $P_{\rm nucl}$.
  At $P_{\rm crit}$ nucleation of the S phase is
  instantaneous, because the energy barrier separating the N
  phase from the S one vanishes.
 }\label{fig:EOS1st}
\end{figure}

\begin{figure}
\includegraphics[width=6.5cm,draft=false]{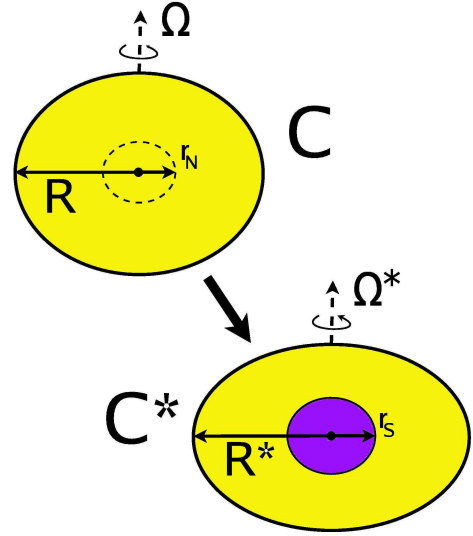}
\caption{Transition from a one-phase configuration ${\cal C}$
with a metastable core of radius $r_{\rm N}$ to a two-phase
configuration ${\cal C}^\star$ with a S-phase core of radius $r_{\rm S}$.
The two configurations have the same baryon number (baryon mass)
and the same angular momentum. From Bejger et al. (2005).
}\label{fig:CCstar}
\end{figure}

First order phase transition allows for metastability of the N phase
at $P>P_0$ (see Fig.\ \ref{fig:EOS1st}). Depending on the magnitude
of the density jump at $P=P_0$ between stable N and S phases, one
has to distinguish two types of 1st order phase transitions.
\parindent 0pt
\vskip 3mm
{\it Weak and moderate 1st order phase
transitions} characterized by  $\rho_{\rm _S}/\rho_{\rm _N}<{3\over
2}+P_0/\rho_{\rm _N}c^2$. A starquake is then triggered by  nucleation of
a droplet of the S phase  in a metastable
core of N phase (see Fig.\ \ref{fig:CCstar}).\par
\begin{figure}
\includegraphics[width=\hsize,draft=false]{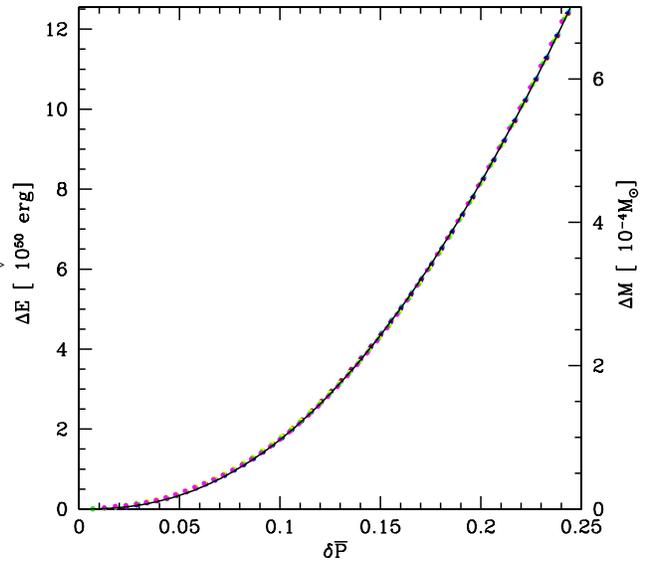}
\caption{(Color online) The energy release due to a corequake of
rotating neutron star as a function of the overpressure $\delta
\overline{P}$ at the center of the metastable N phase core,
for a model of the EOS with a moderate 1st order phase transition.
The points of different color correspond to different values of the
total angular momentum of rotating star, but they all lie on a
same line (Zdunik et al., 2006b,
to be published). Notice, that rotation is very rapid,  with
kinetic/gravitational energy ratio up to $T/W\simeq 0.1$.}
\label{fig:EdeltaP-J-K}
\end{figure}

\vskip 2mm
{\it Strong 1st order phase transitions} characterized by
$\rho_{\rm _S}/\rho_{\rm _N}>{3\over 2}+P_0/\rho_{\rm _N}c^2$.
Configurations with $P_{\rm c}>P_0$ with small
S cores are then {\it unstable} even under full equilibrium conditions
(i.e., for no metastability).
They collapse into configurations
with a large S phase core.
\vskip 3mm
\parindent 21pt
\parindent 21pt
One finds that in all cases $\Delta E$ is, to a very good approximation,
independent of $J$ of
collapsing configuration, provided the energy release is
calculated at fixed overcompression in the center of the
metastable N-star core, $\delta\overline{P}\equiv (P_{\rm c}-P_0)/P_0$.
Generally, a starquake is triggered for $P_{\rm c}=P_{\rm
nucl}$, when the S phase nucleates, which initiates a
transition depicted in  Fig. \ref{fig:CCstar}.
Independence of $\Delta E(\delta\overline{P})$ from $J$ of
collapsing  configuration is clearly seen in Fig.\
\ref{fig:EdeltaP-J-K}. All points lie on the same curve, which
coincides with that obtained for nonrotating case.

Independence of the energy release of $J_{\rm i}$ is valid
for both weak, moderate, and strong 1st order phase transition.
It has a paramount practical importance: simple  1-D
calculations for non-rotating stars
are sufficient to get correct value of the
energy release for  even rapidly rotating
stars.
\section{Conclusions}
\label{sect:Conclusions}

For isolated pulsar without phase transition in the center the evolution
of the period during  spin-down is smooth and monotonic. For
stars with baryon mass below the maximum allowable one for static stars,
$M_{\rm b,max}^{\rm (stat)}$, the stellar angular
momentum $J$ is monotonically decreasing with decreasing rotation frequency
$f$, ${\rm d}J/{\rm d}f>0$. For  supramassive stars, with
fixed baryon mass $M_{\rm b}>M_{\rm b,max}^{\rm (stat)}$, the derivative
${\rm d}J/{\rm d}f$ changes sign very close to the Keplerian (mass shedding)
limit. However, in this region the effect of spin-up by the
angular momentum loss results directly from
effects of general relativity,
and has nothing to do with the equation of state
(Cook et al. 1992, Zdunik et al. 2004). The relevant value of
$f$ is then so close to the mass shedding limit, that the star is
susceptible to many other instabilities and
hence it is unlikely to be  observed as a radio pulsar.

A softening of the EOS of dense matter due to a phase
transition or appearance of hyperons can affect spin evolution
tracks of isolated pulsars. In particular, back bending epochs
and instabilities followed by corequakes may appear. One
parameter families (sequences) of rigidly rotating
configurations loosing angular momentum have a
stability/instability  structure identical with that of
spherical non-rotating family. Energy release associated with
a corequake in rotating pulsar depends only on the
overpressure in the center of its metastable core and does not
depend on the rotation rate of collapsing configuration. This
independence on the initial rotation rate holds universally,
for weak, moderate, and  strong phase transitions,  and with
or without metastability. This means that the energy release
in a corequake in a rotating star can be calculated, with
a very good precision, in 1-D nonrotating model with the same
pressure  at the core center.

It should be mentioned that there is still no direct observational evidence
of the back bending  phenomenon. Although there exist many
isolated pulsars with a measured {\it  decreasing}
 period, they are located in
globular clusters and the effect of $\dot P <0$ is
usually explained by the
acceleration in the globular cluster gravitational field
(for a recent review see Camilo \& Rasio 2005).
The effect of period clustering (which would be another consequence
of the phase-transition  softening of EOS) is not clear from observational
point of view. There  seems to be  some evidence for such effect
(e.g. Chakrabarty 2004) for accreting  neutron stars, but not
for isolated neutron stars considered in the
present paper. However, for accreting neutron
stars  the  effect of back bending is expected to be strongly
suppressed due to the simultaneous increase of the stellar
mass during accretion (Zdunik et al. 2005).
There might be  some indirect evidence for back bending
hided in the evolutionary history of pulsars.
Some evolutionary scenarios applied to the observed pulsars
result in unrealistically short initial period if one assumes
standard magnetic dipole braking. This paradox  could
be explained by the existence of the back-bending epoch during
pulsar's life (Spyrou \& Stergioulas 2002, Zdunik et al. 2006).
Finally, measurements of very large values of braking indices
(Johnston \& Galloway 1999)  could be an
 indication of the closeness to $\dot{f}=0$
state associated with back bending. However, this might be explained
in a more standard way, by the behavior connected
with the glitch phenomena (Johnston \& Galloway 1999,
Alpar \& Baykal 2006).

This work was partially supported by the Polish MNiI grant no.
1P03D-008-27.


\end{document}